\documentclass[runningheads]{llncs}
\usepackage{graphicx}
\usepackage{amsmath,amssymb,multicol,trsym,multirow,bm}
\newcommand{\R}{\mathbb{R}}
\usepackage{url}
\usepackage{pgfplots}

\pgfplotsset{compat=1.17}

\begin{document}
\title{Quantisation Scale-Spaces\thanks{ This work 
		has 
		received funding from the European 
		Research Council  (ERC) under the European Union's Horizon 2020 
		research and 
		innovation programme (grant agreement no. 741215, ERC Advanced 
		Grant INCOVID).}}

\author{Pascal Peter}
\authorrunning{P. Peter}
\institute{Mathematical Image Analysis Group,
	Faculty of Mathematics and Computer Science,\\ Campus E1.7,
	Saarland University, 66041 Saarbr\"ucken, Germany.\\
	peter@mia.uni-saarland.de}
\maketitle              
\begin{abstract}
Recently, sparsification scale-spaces have been obtained as a sequence of 
inpainted images by gradually removing known image data. Thus, these 
scale-spaces rely on spatial sparsity. In the present paper, we show that 
sparsification of the co-domain, the set of admissible grey values, also 
constitutes scale-spaces with induced hierarchical quantisation techniques. 
These quantisation scale-spaces are closely tied to information theoretical 
measures for coding cost, and therefore particularly interesting for 
inpainting-based compression. Based on this observation, we propose a 
sparsification algorithm for the grey-value domain that outperforms uniform 
quantisation as well as classical clustering approaches.

\end{abstract}
\section{Introduction}

Image inpainting \cite{MM98a} reconstructs an image from 
a mask that specifies known pixel data at a subset of all image 
coordinates.  C\'ardenas et al.~\cite{CPW19} have shown that a wide variety of 
inpainting operators can create scale-spaces: Sequentially removing points from 
the mask yields inpainting results that form a scale-space with the mask 
density as a discrete scale dimension. 
The order in which the data is removed, the sparsification 
path, can impact the reconstruction quality significantly.
Implicitly, many inpainting-based compression approaches (e.g. 
\cite{GWWB08,Pe19,SPMEWB14}) and associated mask optimisation 
strategies 
(e.g. \cite{APW17,CRP14,MHWT12}) rely on these sparsification scale-spaces: 
They choose sparse masks as a compact representation of the image and aim for 
an accurate reconstruction from this known data.

However, there is another aspect of compression, that has hitherto not been 
explored from a scale-space perspective: The known data has to be stored, which 
means the information content of the grey or colour values plays an important 
role. In particular, all contemporary compression codecs use some form of 
quantisation combined with variants of entropy coding, no matter if they rely 
on transforms  \cite{PM92,TM02}, inpainting \cite{GWWB08,Pe19,SPMEWB14}, or 
neural networks  \cite{RB17}. Quantisation reduces  the amount of different 
admissible values in the co-domain to lower the Shannon entropy \cite{Sh48a}, 
the measure for information content.

Sparsification scale-spaces have demonstrated that there are
viable scale-space concepts beyond the classical ideas that rely on partial 
differential equations (PDEs) \cite{AGLM93,Ii62,SchW98,We97} or evolutions 
according to pseudodifferential operators \cite{DFGH04,SW16}. In the present 
paper, we show that quantisation methods can also imply scale-spaces and that 
this leads to practical consequences for inpainting-based compression. In 
this application context, quantisation has so far only been the main focus in 
the work of Hoeltgen et al.~\cite{HPB18} who compared multiple different 
strategies. However, they did not consider 
scale-space ideas and concluded that non-uniform quantisation does not offer 
advantages over uniform ones in inpainting-based compression. 

\medskip
{\bf Our contributions.} Our first results are of theoretical nature: We 
propose quantisation scale-spaces based on the broad class of hierarchical 
quantisation methods inspired by Ward clustering \cite{Wa63}: The image is 
gradually simplified by merging level sets. Not only do these sequences of 
quantised images satisfy all essential scale space-properties, we also show 
that our Lyapunov criterion guarantees a decrease of the coding cost.

This observation motivates our practical contributions: We demonstrate 
that committed (i.e. structure-adaptive) quantisation scale-spaces are 
particularly useful. They allow the design of flexible, task-specific 
quantisation approaches. As a concrete application we use the natural ties of 
quantisation scale-spaces to information theory for improved inpainting-based 
compression.

\medskip
{\bf Organisation.} In Section~\ref{sec:quantisation}, we propose quantisation 
scale-spaces and their properties, compare them to 
sparsification scale-spaces in Section \ref{sec:combination}, and discuss 
practical applications in Section~\ref{sec:compression}. 
Section~\ref{sec:conclusion} concludes with a final discussion.

\section{Quantisation Scale-Spaces}
\label{sec:quantisation}

While there is a plethora of quantisation approaches designed for specific 
purposes, not all of them correspond to scale-spaces.
Thus, we first define a suitable class of quantisation 
methods without imposing too many restrictions 
on their design.  To this end, we use a very general 
hierarchical clustering idea inspired by the classical 
algorithm of Ward \cite{Wa63}. Intuitively, these approaches reduce the 
quantisation levels by mapping exactly two of the fine grey values to 
the same coarse quantisation value.
In the following we consider only 1-D vectors, but these represent images of 
arbitrary dimension with $N$ pixels (e.g. a row by row representation in 2-D).

\newpage
\medskip
\noindent
{\bf Definition 1: Hierarchical Quantisation Function}\\[1mm] 
Consider a quantised image $\bm f \in \{v_1,...,v_Q\}^N$ with $N$ pixels and 
$Q$ quantisation levels $v_1 < v_2 < \cdots < v_Q$.
A function 
$q: V:=\{v_1,...,v_Q\} \rightarrow \hat{V} := \{\hat{v}_1, ..., 
\hat{v}_{Q-1}\}$ mapping to coarse quantisation levels $\hat{V}$ with
 $v_1 \leq \hat{v}_1 < \hat{v}_2 < \cdots <\hat{v}_{Q-1} \leq 
v_Q$ 
is a \emph{hierarchical quantisation} function if it fulfils
\begin{align}
\begin{split}
\exists \, s,t \in \{1,...,Q\}, \, s\neq t: \, \forall i,j \in \{1,...,Q\}, \, 
i 
\neq j 
: \\ 
q(v_i) = q(v_j) \, \, \Longleftrightarrow \, \, i,j \in 
\{s,t\} \, .
\label{eq:hierarchical}
\end{split}
\end{align}
\medskip

Eq.~\eqref{eq:hierarchical} guarantees that exactly $v_s$ and $v_t$ from the 
fine grey value range $V$ are mapped to the same value $v_r := q(v_s)$ from the 
coarse range $\hat{V}$. Since $\hat{V}$ contains exactly one value less, 
this simultaneously implies a one-to-one mapping for all remaining $Q-2$ values 
in $V \setminus \{v_s, v_t\}$ and $\hat{V} \setminus \{v_r\}$. 

For our purposes we also consider quantisation as an 
operation on level sets. 
Let a level set $L_k$ of the image $\bm{f}=(f_i)_{i=1}^N$ be given by the 
locations with grey value $v_k$:\begin{equation}
L_k = \{  i  \, | \, f_i = v_k  \}, \quad k \in \{1,...,Q\} \, .
\label{eq:levelset}
\end{equation}
With this notation, our definition of a hierarchical quantisation step is 
equivalent to merging exactly the two level sets $L_s$ and $L_t$, while leaving 
the rest untouched.

Note that we do not specify which level sets need to be merged. Moreover, the 
only restriction that we impose to the target range $\hat{V}$ is that it does 
not exceed the original grey level range. As we demonstrate in 
Section~\ref{sec:compression}, hierarchical  quantisation covers a very broad 
range of quantisation methods that can be tailored to specific applications.
With the formal notion of hierarchical quantisation functions we are now 
suitably equipped to define quantisation scale-spaces.

\medskip
\noindent
{\bf Definition 2: Quantisation Scale-Spaces}\\[1mm] 
Consider the original image $\bm{f}=(f_i)_{i=1}^N \in V^0:=\{v_1,...,v_Q\}^N$, 
$v_1 < v_2 < \cdots < v_Q$ containing $Q$ initial grey levels. Let $\bm f^0 := 
\bm 
f$ and consider hierarchical quantisation functions $q^\ell$ 
with $\ell \in \{1,...,Q-1\}$. Each function $q^\ell$ maps 
$V^{\ell-1}$ with $|V^{\ell-1}|=Q-\ell+1$ to a quantised range $V^{\ell}$ with 
$Q-\ell$ 
grey values. 
The quantisation scale-space is then given 
by the family of images
\begin{align}
\bm f^0 &= \bm f \, , \\
\bm f^\ell &= \left(q^\ell 
\left(f^{\ell-1}_1\right), ..., q^\ell 
\left(f^{\ell-1}_{N}\right)\right), \quad \ell \in \{1,...,Q-1\} \, ,
\label{eq:sss}
\end{align}
with discrete scale parameter $\ell \in \{0,...,Q-1\}$.

\medskip
Since we use hierarchical quantisation functions in Definition~2, we make sure 
that the range of admissible grey values is reduced by one when we transition 
from scale $\ell$ to scale $\ell+1$ with $q^{\ell+1}$. Note that we do not 
require that $\bm f$ 
actually contains all values of the range $V^0$. Thus, some of its level sets 
might be empty. For 8-bit images, we have $Q=256$.

\newpage
\subsection{Scale-Space Properties}

In the following we show that quantisation scale-spaces in the 
sense of Definition~2 fulfil six essential scale-space properties 
\cite{AGLM93}. For some of 
these properties, it is helpful to argue on level sets. Therefore, we introduce 
the notation $L^\ell_k$ for the level set of the value $v^\ell_k$ (as in 
Eq.~\eqref{eq:levelset}) on scale $\ell$.


\medskip
\noindent
{\bf Property 1: Original Image as Initial State}\\[1mm]
For $\ell=0$, Definition 2 directly yields $\bm f^0 = \bm f$ which can be seen
as quantisation with the identity function.

%
\medskip
\noindent
{\bf Property 2: Semigroup Property}\\[1mm] 
Due to Definition~2, we can construct the scale-space in a cascadic manner 
by subsequent merging of level sets. In particular, the hierarchical 
quantisation criterion from Eq.~\eqref{eq:hierarchical} implies that 
$q^{\ell+1}$ 
merges exactly two level sets $L^\ell_s$ and $L^\ell_t$ into a new level set 
$L^{\ell+1}_r$, while the rest remains unchanged.
Therefore, a quantised image $\bm{f}^{\ell+n}$ can be obtained from $\bm f^0$ 
in $\ell + n$ merging steps or, equivalently, from $\bm f^\ell$ in $n$ merging 
steps.

\medskip
\noindent
{\bf Property 3: Maximum--Minimum Principle}\\[1mm]
According to Definition~1, hierarchical quantisation functions do not
expand the grey value range. Thus, all quantised values lie in the range 
$[v_1,v_Q]$ and the maximum--minimum principle is fulfilled.


\medskip
\noindent
{\bf Property 4: Lyapunov Sequences}\\[1mm] 
Lyapunov sequences play a vital role in describing the sequential simplification
of the image with increasing scale parameter \cite{We97}. Multiple different 
Lyapunov sequences can be specified that emphasise specific types of 
simplification. In analogy to sparsification scale spaces 
\cite{CPW19}, for $J=\{1,...,N\}$, we can use the total image contrast 
\begin{equation}
V(\bm{f}^\ell) \;:=\; \max_{i \in J} f_i^\ell - \min_{i \in J} f_i^\ell
\end{equation} to define a Lyapunov sequence based on the maximum--minimum 
principle. Thus, the image is visually simplified by shrinking the contrast 
successively.

However, for progressive quantisation, it is more meaningful to consider a 
different Lyapunov sequence that directly reflects the intention behind 
quantisation in compression applications: the entropy as a measure of 
information content and storage cost. For the Shannon entropy \cite{Sh48a}, we 
need the probabilities of the grey values occurring in the image $\bm f^\ell$. 
We can define them via 
the level sets as 
\[
p^\ell_k := \frac{|L^\ell_k|}{N} \, ,
\]
where $|\cdot|$ denotes the cardinality. Then, the binary entropy $H$, 
with $\log_2$ denoting the logarithm to the base 2, is given by
\begin{equation}
H(\bm f^\ell) = - \sum_{\substack{\large k=1, \\ p_k \neq 0}}^{Q-\ell} p_k 
\log_2 
p_k \, . \label{eq:entropy} 
\end{equation}
It constitutes the minimal average bit cost of storing a grey value of the image
$\bm f^\ell$ that can be achieved by any binary zeroth order entropy coder. 
Since Definition~2 allows empty level sets, these are excluded from the sum in 
Eq.~\eqref{eq:entropy}.

 Note that this entropy criterion for simplification 
differs significantly from the notion of entropy used in PDE-based scale-spaces 
\cite{We97}: There, the entropy is defined directly on the grey values instead 
of their histogram and thus does not reflect storage cost. In the following, we 
prove that the Shannon entropy constitutes a Lyapunov sequence.

\bigskip
\noindent
{\bf Proposition 1: Shannon Entropy constitutes a Lyapunov Sequence}\\[1mm] 
For a quantisation scale-space according to Definition~2, the Shannon entropy 
decreases monotonically with the scale parameter, i.e. for all $\ell$
\begin{equation}
H(\bm f^\ell) \geq H(\bm f^{\ell+1}) \, .
\label{eq:lyapunov}
\end{equation}

\begin{proof}
Let $L^\ell_i$, $L^\ell_j$ denote the two level sets that are merged to obtain 
$L^{\ell+1}_r = L^\ell_i \cup L^\ell_j$ according to 
Eq.~\eqref{eq:hierarchical}.
The 
probabilities $p_i$ and $p_j$ are additive under merging, i.e.
\begin{equation}
p^{\ell+1}_r =  \frac{|L^{\ell+1}_r|}{N} =  \frac{|L^\ell_i|}{N} +  
\frac{|L^\ell_j|}{N} = p^\ell_i + p^\ell_j \, .
\end{equation}
This implies equality in Eq.~\eqref{eq:lyapunov} if $L^\ell_i$ or 
$L^\ell_j$ are empty. Assuming that both level sets are non-empty, the entropy 
is strictly decreasing for 
increasing scale parameter $\ell$:
\begin{align}
H(\bm f^{\ell+1}) &\;=\; - p^{\ell+1}_r \log_2 p^{\ell+1}_r - \sum_{k=1, k \neq 
	r}^{Q-\ell} p^{\ell+1}_k \log_2 p^{\ell+1}_k  \\
&\;=\; - (p^\ell_i + p^\ell_j) \log_2 
(p^\ell_i + p^\ell_j) - \sum_{k=1, k \notin \{i,j\}}^{Q-\ell} p^\ell_k \log_2 
p^\ell_k \\
&\;<\; - p^\ell_i \log_2 
(p^\ell_i) - p^\ell_j \log_2 
(p^\ell_j) - \sum_{k=1, k \notin \{i,j\}}^{Q-\ell} p^\ell_k \log_2 
p^\ell_k = H(\bm f^\ell)
\end{align}
Note that in the first step of the computation above we have used that there is 
a one-to-one mapping between all coarse and fine level sets not affected by the 
merge. The last step  follows from the fact that $p^\ell_i > 
0$, $p^\ell_j > 0$, and $\log_2$ is monotonically increasing, hence
\begin{align}
(p^\ell_i + p^\ell_j)\log_2 
(p^\ell_i + p^\ell_j)
\;>\; p^\ell_i \log_2 
(p^\ell_i) + p^\ell_j \log_2 
(p^\ell_j) \, .
\end{align} 
Thus, we have shown that each step in a quantisation scale-space does not 
increase information content and, if no level sets are empty, it even strictly 
decreases the entropy.
\hfill$\qed$
\end{proof}

\newpage
\medskip
\noindent
{\bf Property 5 (Invariances)}\\[1mm]
A quantisation according to Definition~1 is a point operation, i.e. it acts 
independently of the spatial configuration of image pixels. Thus, it is 
invariant under any permutation of the image pixels. Moreover, it is invariant 
under any brightness rescaling operations that keep the histogram, and thus the 
level sets, intact.


\medskip
\noindent
{\bf Property 6 (Convergence to a Flat Steady-State)}\\[1mm]
According to Definition~2, the final image $\bm{f}^{Q-1}$ contains $Q-(Q-1)=1$ 
grey value and is thus flat.

\medskip
Now that the important properties of quantisation scale-spaces have been 
verified, we briefly compare them to their spatial relatives, the 
sparsification scale-spaces \cite{CPW19}. Then, in Section 
\ref{sec:compression}, we also consider practical applications for this novel 
type of scale-space.


\setlength{\fboxsep}{0pt}
\begin{figure*}[tph]
	\tabcolsep=1pt
	\begin{tabular}{ccccc}
		& $100\,\%$  & $\approx 64\,\%$ &$\approx 8\,\%$ & $\approx 1\,\%$ \\
		\rotatebox{90}{\hspace{0.5em}\small inpainting mask} &
		\fbox{%
			\includegraphics[width=0.228\linewidth]{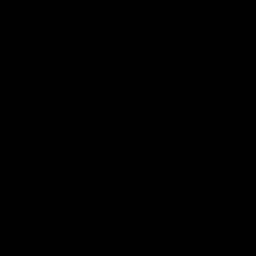}%
		}&
		\fbox{%
			\includegraphics[width=0.228\linewidth]{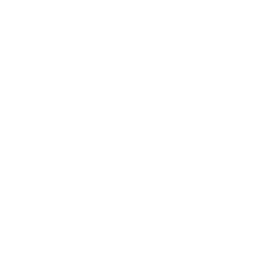}%
		}&
		\fbox{%
			\includegraphics[width=0.228\linewidth]{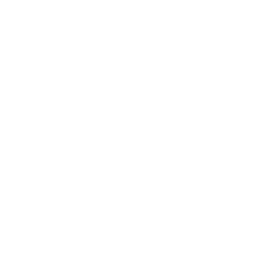}%
		}&
		\fbox{%
			\includegraphics[width=0.228\linewidth]{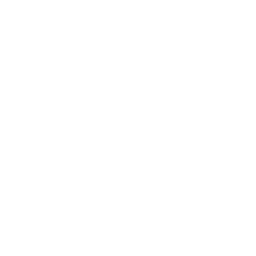}%
		}\\
		\rotatebox{90}{\hspace{2.5em}\small $q=64$} &
		\includegraphics[width=0.228\linewidth]{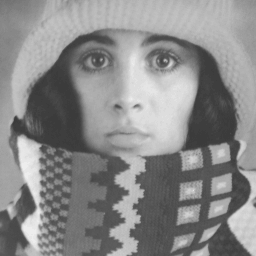}&
		\includegraphics[width=0.228\linewidth]{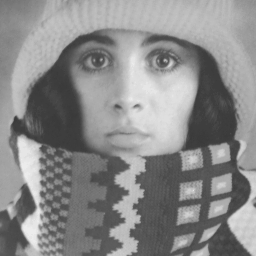}&
		\includegraphics[width=0.228\linewidth]{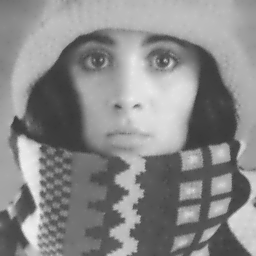}
		&
		\includegraphics[width=0.228\linewidth]{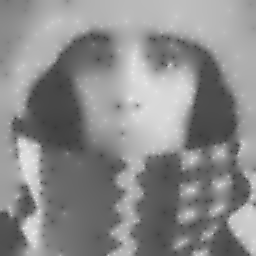}\\
		\rotatebox{90}{\hspace{2.5em}\small $q=16$} &
		\includegraphics[width=0.228\linewidth]{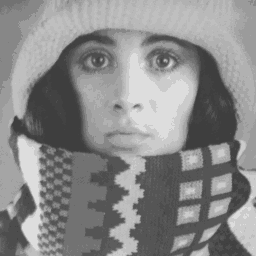}%
		&
		\includegraphics[width=0.228\linewidth]{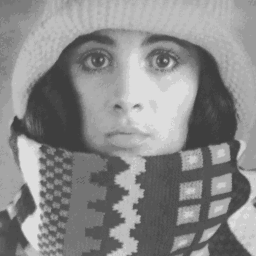}&
		\includegraphics[width=0.228\linewidth]{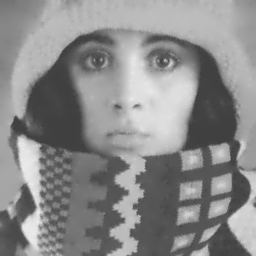}%
		&
		\includegraphics[width=0.228\linewidth]{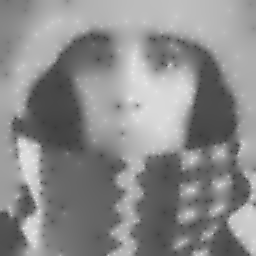}\\
		\rotatebox{90}{\hspace{2.5em}\small $q=4$} &
		\includegraphics[width=0.228\linewidth]{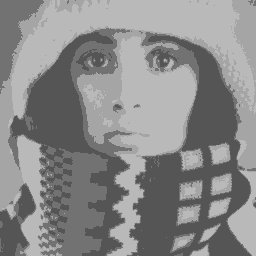}
		&
		\includegraphics[width=0.228\linewidth]{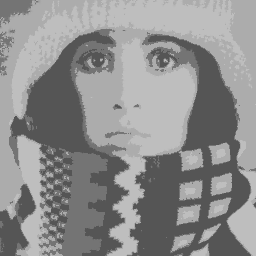}&
		\includegraphics[width=0.228\linewidth]{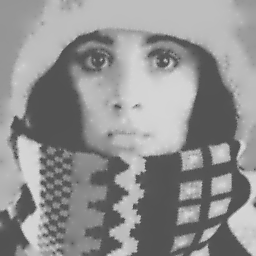}
		&
		\includegraphics[width=0.228\linewidth]{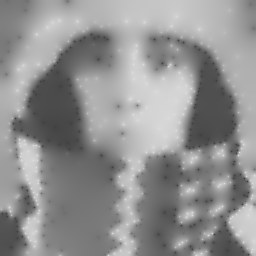}\\
		\rotatebox{90}{\hspace{2.5em}\small $q=1$} &
		\includegraphics[width=0.228\linewidth]{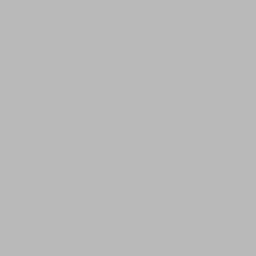}%
		&
		\includegraphics[width=0.228\linewidth]{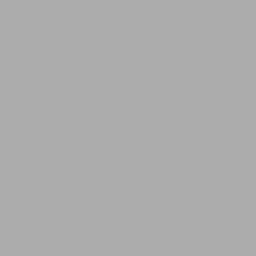}&
		\includegraphics[width=0.228\linewidth]{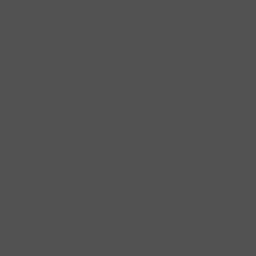}
		&
		\includegraphics[width=0.228\linewidth]{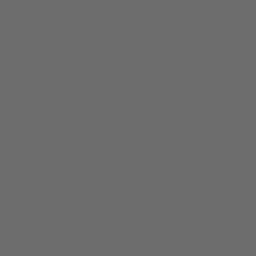}\\
	\end{tabular}
	\caption{\label{fig:scaledimensions} \textbf{Sparsification 
			versus Quantisation Scale-Spaces}: Both the reduction of known data 
		(from left 
		to right) and the reduction of grey values (from top to bottom) 
		simplifies the test image \emph{trui}. Here, we used probabilistic 
		sparsification 
		\cite{CPW19,MHWT12} for  the 
		spatial domain and our quantisation by sparsification for the grey 
		level domain. Interestingly, for very low 
		quantisation levels $q$, results might be visually more pleasant for 
		lower mask densities, compare e.g. 8\,\% to	64\,\% 
		at 4 grey levels. This impression results from the smooth interpolation 
		by 	diffusion, that avoids unpleasant discontinuities and simultaneously
		expands the number of grey levels in the reconstructed image beyond 
		those of the mask.}
\end{figure*}

\section{Relations to Sparsification Scale-Spaces}
\label{sec:combination}

Sparsification scale-spaces \cite{CPW19} rely on a series of nested inpainting 
masks that specify known data for an image with domain $\Omega$ with 
$|\Omega|=N$. They start with a full mask $K^0 = \Omega$ and successively reduce
$K^\ell$ to $K^{\ell+1}$ by removing exactly one pixel coordinate from this 
set. From each of these sets, an image $\bm u^\ell$ can be inpainted by solving
\begin{equation}
\bm{C}^\ell \! \left(\bm{u}^\ell-\bm{f}\right) \;-\; 
\left(\bm{I}-\bm{C}^\ell\right)  \bm{A}\left(\bm{u}^\ell\right) 
\bm{u}^\ell \;=\; \bm{0} \qquad (\ell=0,...,N\!-\!1).
\label{eq:inpainting}
\end{equation}
Here, $\bm C^\ell \in \R^{N \times N}$ is a diagonal matrix that corresponds to 
the known data set $K^\ell$, and $\bm A \in \R^{N \times N}$ implements an 
inpainting operator including boundary conditions. While there are many viable 
choices for $\bm A$, we only consider homogeneous diffusion inpainting 
\cite{Ii62} here, where $\bm A$ corresponds to a finite difference 
discretisation of the Laplacian with reflecting boundary conditions. For more 
details on sparsification 
scale-spaces we refer to C\'ardenas et al.~\cite{CPW19}.

Both sparsification and quantisation scale-spaces have a discrete scale 
parameter. The corresponding amount of steps is determined by the image 
resolution in the spatial setting and by the grey value range in the 
quantisation setting. Additionally, sparsification scale-spaces do not have an 
ill-posed direction: For a known order of pixel removals, one can easily go 
from coarse scales to fine scales.  Due to the many-to-one mapping by merging 
level sets, this is not the case for quantisation scale-spaces.
In compression applications, this is no issue, since the original image is 
available. However, one cannot go to finer quantisations, i.e. from an 8-bit to 
a high dynamic range image.


Interestingly, we can combine sparsification and quantisation scale-spaces. For 
an image $\bm f \in V^N$ with $|V|=Q$, consider the sparsification path given 
by $(K^\ell)_{\ell=0}^{N-1}$. Now, we can 
define a quantisation scale-space on $K^\ell = \{i_1,...,i_n\}$ with 
$n:=N-\ell$ known 
pixels. Thereby, following 
Definition~2, we consider the known grey values $\bm g^\ell := (f_{i_1}, ..., 
f_{i_n})$ instead of the full image $\bm f$. Quantisation with scale 
parameter 
$m$ yields $n$ pixels from $V^m$ with 
$q:=|V^m|=Q-m$ quantisation values. 
The reconstruction $\bm 
u^{\ell,m}$ now depends on the respective scale parameters $\ell$ of the 
sparsification and $m$ of the quantisation scale-space.

This does not affect our theoretical results, since none of the 
properties relies on the spatial configuration of the image pixels. If an 
inpainting operator fulfils the maximum--minimum principle, this carries over 
to 
the inpainted image, and thereby also the total contrast of the known data. We 
can thus even extend Properties 3 and 4 to the reconstructions $\bm 
u^{\ell,m}$.

An investigation of both scale dimensions in Fig.~\ref{fig:scaledimensions} 
yields surprising results: A lower amount of known data at the same 
quantisation scale 
can yield visually more pleasing results (e.g. $q=4$ and $64\,\%$ vs. $8\,\%$ 
mask 
density). This results from the fact that smooth inpainting can fill in 
additional grey levels, hence the inpainted areas are less coarsely quantised 
than the known data.

Finally, both types of scale-spaces can be committed to the image, i.e. adapted 
to the image structure. In the following section, we show that this allows the 
design of highly task-specific quantisation approaches.

\section{Applications to Quantisation and Compression}
\label{sec:compression}

The results of practical experiments depend significantly on the order in which
known data are removed and level sets are merged. For the spatial scale-space, 
we use an adaptive sparsification path obtained with the algorithm from 
\cite{CPW19}. First, we propose several committed and uncommitted 
quantisations. We show some exemplary results on the test image \emph{trui} 
that has been also used in \cite{CPW19}.


\setlength{\fboxsep}{0pt}
\begin{figure*}[tph]
	\tabcolsep=2.5pt
	\begin{tabular}{cccc}
		& uniform & \textbf{our sparsification} & Ward \\
		\rotatebox{90}{\hspace{4em}\small $q=16$} &
		\includegraphics[width=0.3\linewidth]{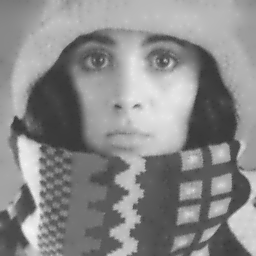}&
		\includegraphics[width=0.3\linewidth]{imgquant/pyrspars-8-16}
		&
		\includegraphics[width=0.3\linewidth]{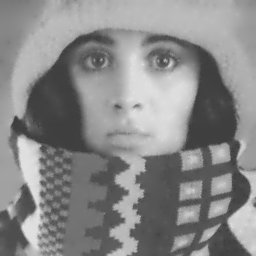}\\
		\rotatebox{90}{\hspace{4.5em}\small $q=8$} &
		\includegraphics[width=0.3\linewidth]{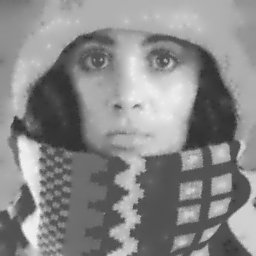}&
		\includegraphics[width=0.3\linewidth]{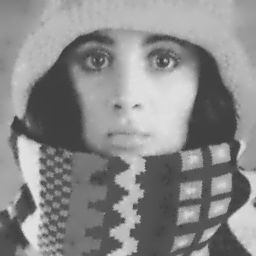}%
		&
		\includegraphics[width=0.3\linewidth]{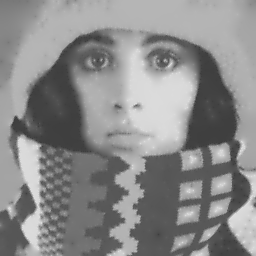}\\
		\rotatebox{90}{\hspace{4.5em}\small $q=4$} &
		\includegraphics[width=0.3\linewidth]{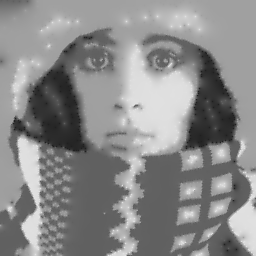}
		&
		\includegraphics[width=0.3\linewidth]{imgquant/pyrspars-8-4}
		&
		\includegraphics[width=0.3\linewidth]{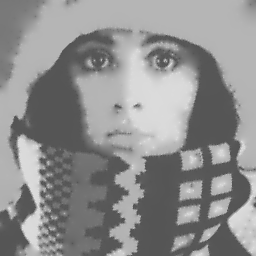}\\
		\rotatebox{90}{\hspace{4.5em}\small $q=2$} &
		\includegraphics[width=0.3\linewidth]{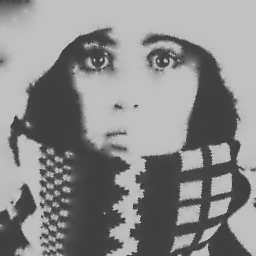}&
		\includegraphics[width=0.3\linewidth]{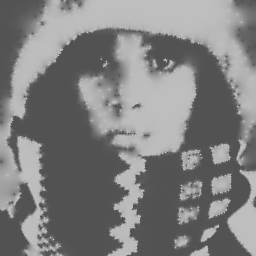}
		&
		\includegraphics[width=0.3\linewidth]{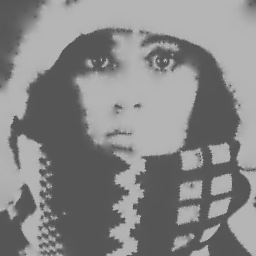}\\
	\end{tabular}
	\caption{\label{fig:comparison} \textbf{Visual Comparison of Quantisation 
			Scale-Spaces}: The uncommitted uniform quantisation scale-space 
			does not preserve details of \emph{trui} as well as the committed 
			sparsification 
			and Ward approaches for low amounts of quantisation levels $q$. The 
			quantisation by sparsification scale-space preserves structures in 
			the facial area, and the hat even down to only two 	quantisation 
			levels (mouth, nose). The Ward method preserves a few more 
		patterns on the left side of the scarf, but looses detail in all other 
		areas.}
\end{figure*}


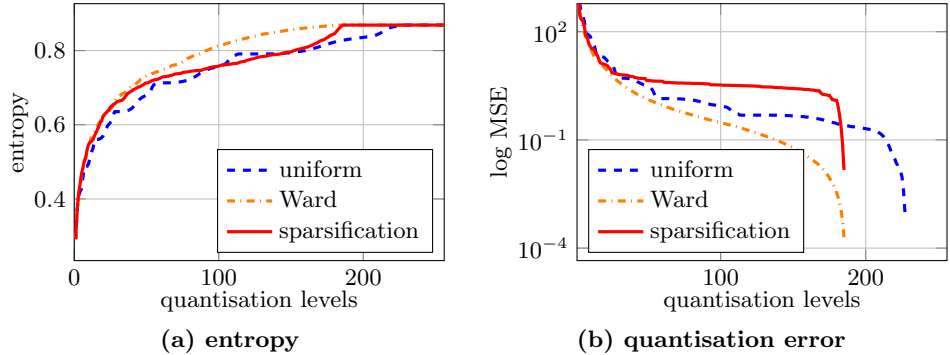
\begin{figure}[ht!]
		\begin{minipage}{0.5\textwidth}	
		\begin{tikzpicture}
		\begin{axis}[height=5cm, width=6.5cm,
		xmin=0, xmax = 256,
		label style={font=\footnotesize},
		tick label style={font=\footnotesize},  
		xlabel={quantisation levels},
		ylabel={entropy},
		ylabel style={yshift=-0.14cm},
		xlabel style={yshift=0.14cm},
		grid = major,
		legend entries={uniform, Ward, sparsification},
		legend cell align=left,
		legend style={legend pos=south east,font=\footnotesize}],
		\addplot+[mark=none, dashed, very thick, blue] table[x index=1, y 
		index=4]
		{data/pyruni8_log.txt};
		\addplot+[mark=none,dashdotted, very thick, orange] table[x index=1, y 
		index=4] {data/ward8_log.txt};
		\addplot+[mark=none, very thick,red] table[x index=1, y index=4] 
		{data/pyrspars8_log.txt};
		\end{axis}
		\end{tikzpicture}\\[-7mm]	
		\begin{center}
			\textbf{(a) entropy}
		\end{center}
	\end{minipage}\hspace{3mm}%
	\begin{minipage}{0.5\textwidth}
		\begin{tikzpicture}
		\begin{axis}[height=5cm, width=6.5cm,
		ymax=600,
		xmin=1, xmax=256,
		ymode=log,
		label style={font=\footnotesize},
		tick label style={font=\footnotesize},  
		xlabel={quantisation levels},
		ylabel={log MSE},
		ylabel style={yshift=-0.14cm},
		xlabel style={yshift=0.14cm},
		grid = major,
		legend entries={uniform, Ward, sparsification},
		legend cell align=left,
		legend style={legend pos=south west,font=\footnotesize}],
		\addplot+[mark=none, dashed, very thick, blue] table[x index=1, y 
		index=3] 
		{data/pyruni8_log.txt};
		\addplot+[mark=none,dashdotted, very thick, orange] table[x index=1, y 
		index=3] 
		{data/ward8_log.txt};
		\addplot+[mark=none, very thick,red] table[x index=1, y index=3] 
		{data/pyrspars8_log.txt};
		\end{axis}
		\end{tikzpicture}\\[-7mm]	
		\begin{center}
			\textbf{(b) quantisation error}
		\end{center}
	\end{minipage}
	\caption{\label{fig:quantgraphs}\textbf{Comparing Scale-Space Quantisation 
	Operators on \emph{trui}.} \textbf{(a)} All three operators guarantee 
	decreasing entropy 
	for less quantisation levels. \textbf{(b)} According to design, Ward 
	clustering achieves the best quantisation error. Due to the logarithmic MSE 
	scale, the graphs are cut off at the point where the error reaches zero. }
\end{figure}

\subsection{Uncommitted and Committed Quantisation}

{\bf Uncommitted Uniform Quantisation:} Hierarchical uniform quantisation can 
be seen as a 
pyramidal approach. On a level with $a:=2^k$ level sets, we have $b:=2^{k-1}$ 
pairs of neighbouring level sets. We progress to $b$ grey values by merging 
$L_{2i-1}$ 
and $L_{2i}$ for $i=1,...,b$ in $b$ steps. A bin containing the minimum grey 
value $v_{\min}$ and maximum value $v_{\max}$ is associated to the new value 
$v_{\min} + \frac 1 2 (v_{\max} - v_{\min})$. We round the intermediate results 
to the next integer.

\medskip
\noindent
{\bf Committed Ward Clustering:} While the method of Ward \cite{Wa63} describes 
a general 
clustering approach, not a quantisation technique, it can be used easily as 
such by choosing the mean squared error (MSE)  on the quantised data as an 
optimality criterion. It 
simply merges two level sets that minimise this criterion. 

\medskip
\noindent
{\bf Committed Quantisation by Sparsification:} Inspired by the spatial 
sparsification 
for adaptive scale-spaces \cite{APW17,CPW19}, we successively remove the one 
grey level that has the lowest impact on the global inpainting error. This 
merging strategy can equivalently be interpreted as an inpainting-based merging 
criterion for Ward clustering that is designed for cases where parts of the 
image are unknown. If all pixels are known this corresponds to regular Ward 
clustering.

\medskip
For both committed quantisations, we assign the 
value of the corresponding coarse level set with the largest histogram 
occurrence to the newly merged set. 

Experiments on the test image \emph{trui} with 8\% known data in 
Fig.~\ref{fig:quantgraphs}(a) verify that the entropy monotonically 
decreases with increasing scale for all three approaches. However, this is 
quite irregular for uniform quantisation, since it does not respect the actual 
distribution of grey levels in the image. The non-uniform methods do not yield 
any changes until they reach $186$ grey levels, since this is the number 
actually occurring in the image. Sparsification leads to a slightly quicker 
descent in entropy, but for coarse scales, all methods yield similar results.

The quantisation error in Fig.~\ref{fig:quantgraphs}(b) shows Ward clustering 
as a clear winner: Unsurprisingly, it yields much better results than uniform 
quantisation due to its adaptivity. Interestingly, sparsification exhibits the 
worst results for medium quantisation levels. The next section reveals 
that this is the intended result of its merging criterion and not a design flaw.

\subsection{Inpainting-based Compression}

The combination of sparsification and quantisation scale-spaces from Section 
\ref{sec:combination} describes the natural rate-distortion optimisation problem
in image compression: We want to find the optimal scale 
parameters $\tilde \ell$ and $\tilde m$ such that for a given storage budget 
$t$, the 
MSE is minimised. If $B(K^\ell, \bm g^{l,m})$ denotes the 
coding cost of the known data, image compression can now be seen as a 
minimisation problem
\begin{equation}
(\tilde \ell, \tilde m) = \underset{(\ell,m)}{\operatorname{argmin}} \| \bm f - 
\bm 
u^{\ell,m} 
\|^2, \quad \textnormal{s.t.} \, \; B(K^\ell, \bm g^{l,m}) < t\,
\label{eq:ratedistortion}
\end{equation}
with  Euclidean norm $\|\cdot\|$  and cost threshold $t$. Here, we 
consider the entropy $H(\bm g^{l,m})$   of the known 
grey values for the budget $B$. We can neglect the 
coding cost of the known data positions since they are identical for all 
our quantisation methods. 
However, we still need to consider overhead. For uniform quantisation, this is 
only the number of grey levels (8 bit). Non-uniform methods also need to store 
the values in $V^m$, which comes down to $- q \log_2(q)$ for 
$q=256-m$ grey 
levels.

\begin{figure}[t]
\begin{minipage}{0.5\textwidth}
	\begin{tikzpicture}
	\begin{axis}[height=5.5cm, width=6.5cm,
	xmin=1, xmax = 256,
	label style={font=\footnotesize},
	tick label style={font=\footnotesize},  
	xlabel={quantisation levels},
	ylabel={total overhead (bit)},
	ylabel style={yshift=-0.14cm},
	xlabel style={yshift=0.14cm},
	grid = major,
	legend entries={uniform, non-uniform},
	legend cell align=left,
	legend style={legend pos=north west}],
	\addplot+[mark=none,dashdotted, very thick, blue] table[x index=1, y 
	index=7] 
	{data/pyruni8_log.txt};
	\addplot+[mark=none, very thick,red] table[x 
	index=1, y index=7] {data/pyrspars8_log.txt};
	\end{axis}
\end{tikzpicture}\\[-7mm]
\begin{center}
	\textbf{(a) overhead} 
\end{center}
\end{minipage}\hspace{3mm}%
\begin{minipage}{0.5\textwidth}	
	\begin{tikzpicture}
\begin{axis}[height=5.5cm, width=6.5cm,
	xmin=10, xmax = 500,
	ymin=0, ymax=45,
	label style={font=\footnotesize},
	tick label style={font=\footnotesize},  
	xlabel={compression ratio},
	ylabel={MSE},
	ylabel style={yshift=-0.14cm},
	xlabel style={yshift=0.14cm},
	grid = major,
	legend entries={uniform, Ward, sparsification},
	legend cell align=left,
	legend style={legend pos=south east}],
	\addplot+[mark=none, dashed, very thick, blue] table[x index=0, y 
	index=1] 
	{data/pyruni.txt};
	\addplot+[mark=none,dashdotted, very thick, orange] table[x index=0, y 
	index=1] 
	{data/ward.txt};
	\addplot+[mark=none, very thick,red] table[x 
	index=0, y index=1] {data/pyrspars.txt};
	\end{axis}
	\end{tikzpicture}\\[-7mm]	
	\begin{center}
		\textbf{(b) compression performance} 
	\end{center}
\end{minipage}
	\caption{\label{fig:compression}\textbf{Compression with Quantisation 
	Scale-Spaces on \emph{trui}.} \textbf{(a)} Non-uniform quantisation creates 
	much higher overhead than uniform approaches. \textbf{(b)} Despite the 
	overhead, quantisation by sparsification yields the best compression 
	performance, while Ward clustering falls behind uniform quantisation. }
\end{figure}
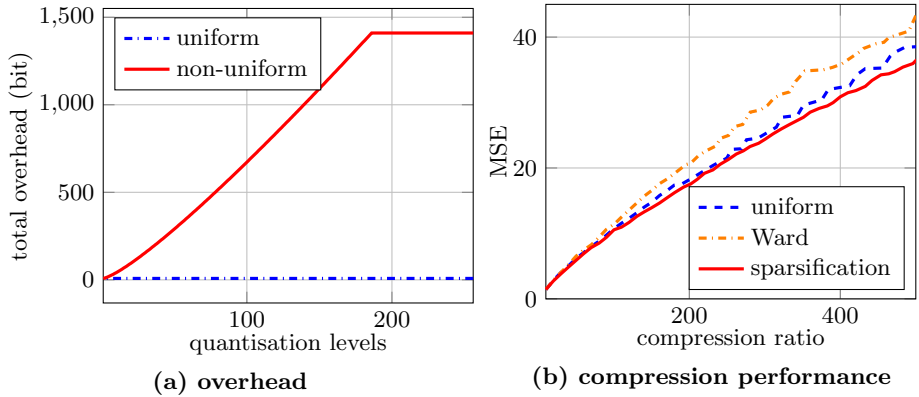

Fig.~\ref{fig:compression}(a) shows that the overhead of non-uniform 
quantisation can easily exceed uniform overhead by a factor 190. 
Consequentially, this is a game changer: In Fig.~\ref{fig:compression}(b) Ward 
clustering is consistently worse for actual compression than a uniform 
quantisation. These findings are consistent with those of 
Hoeltgen~et~al.~\cite{HPB18} who also tested non-scale-space quantisations such 
as k-means 
clustering~\cite{Ll82}. We did not consider those here since they do not offer 
better results.

However, our quantisation by sparsification does not only beat Ward clustering 
by more than 20\%, but also uniform quantisation by more than 10\%. This is a 
direct result of the correct error measure for compression during the choice of 
the quantisation path. Minimising the MSE on the inpainted image is the true 
goal of this application, while Ward clustering only ensures accurate known 
values. Thus, committed quantisation scale-spaces can be useful to adapt to 
different applications.


\section{Conclusions}
\label{sec:conclusion}

We have established quantisation scale-spaces as the co-domain counterpart
of spatial sparsification scale-spaces: They act on the grey value domain 
instead of the pixel domain. Both of these approaches successively sparsify 
their corresponding discrete domains and we can even combine them in a 
meaningful way for concrete applications in compression. 

However, there are also notable differences: In contrast to spatial 
sparsification, the Lyapunov criterion for quantisation scale-spaces simplifies 
the known data set in terms of its Shannon entropy, thus yielding a guaranteed 
reduction of the coding cost. From the viewpoint of inpainting-based image 
compression, we can now interpret rate-distortion optimisation as the selection 
of suitable scale parameters in both scale-spaces.

In future research, we are going to consider quantisation scale-spaces on 
colour spaces and further investigate their impact on compression in a more 
practical setting, including a larger scale evaluation. 

%
%
%
 \bibliographystyle{splncs04}
 \bibliography{bib}

\end{document}